\documentstyle[12pt,epsf,epsfig,wrapfig]{article}  
\begin{document}  
\newtheorem{thm}{Theorem}
\newtheorem{cor}{Corollary}
\newtheorem{Def}{Definition}
\newtheorem{lem}{Lemma}
\begin{center}  
{\large \bf  
Clebsch-Gordan coefficients and the binomial distribution.
\ }  
\vspace{5mm}  

Paul O'Hara
\\  
\vspace{5mm}  
{\small\it  
Dept. of Mathematics, Northeastern Illinois University, 5500  
North St. Louis Avenue, Chicago, IL 60625-4699, USA. email:
pohara@neiu.edu \\}  
\end{center}  
\begin{abstract} A class of Clebsch-Gordan coefficients are
derived from the properties of conditional probability using the
binomial distribution. In particular,
in the case of $l=l_1+l_2$ it is shown that
$$\left<l_1/2-k_1, l_2/2-k_2|l/2, k=k_1+k_2\right>^2
=\frac{\left(\begin{array}{c} 
l_1\\k_1\end{array}\right)
\left(\begin{array}{c}l_2\\k_2\end{array}\right)}{\left(\begin{array}{c}l\\k
\end{array}\right)}$$ 
Pacs: 3.65, 2.50.Cw.    
\end{abstract}

\section {Introduction}

There appear to be two standard ways of calculation C-G coefficients in 
quantum mechanics. One method is to combine the ``ladder'' operator 
approach with 
orthogonality conditions. The second method is to use some type of closed form
\cite{brink} which allows the coefficients to be calculated directly. 
However, the latter approach is at times considered ``tedious'' 
\cite{brink} and does not reveal
any new information into the nature of these coefficients.

In this paper we prove a theorem which not only permits a special class of
C-G coefficients to be calculated from a simple formula but also
directly connects them to both the hypergeometric and binomial
distributions of classical probability. 

Before formulating and proving the theorem, we first define some notation.
Let $L=(L_1, L_2, L_3)$ denote the
angular momentum operator and define
\begin{eqnarray} L^{\pm}=L_x\pm iL_y.
\end{eqnarray}
Then
\begin{eqnarray}
L^2 &=& L^2_x + L^2_y + L^2_z\\
&=& L^{-}L^{+}+L^2_z+L_z,
\end{eqnarray}
from which it follows that
\begin{eqnarray} L^2\left|l, m\right>=l(l+1)\left|l, m\right>,
\end{eqnarray}
and 
\begin{eqnarray} L^{\pm}\left|l, m \right>=[(l\mp m)(l\pm m+1)]^{1/2}
\left|l, m\pm 1\right>,
\end{eqnarray}
where $\left|l, m\right>$ is an eigenvector of $L^2$ and $L_z$.  
Similarly, the basis vectors\newline
$\left|l, m\right>\dots \left|l-n, m\right>$ are eigenvectors of 
$L^2$ and $L_z$ with
$-l\le m \le l$.
Now consider the operator  ${\bf L}=L_1+L_2$, $L_1$
and $L_2$ being angular momentum operators as defined above. Denote the basis 
vectors of
${\bf L}^2, {\bf L}_z$ by $\left|LM\right>$,
where $|l_1-l_2|\le
L\le |l_1+l_2|$ and denote the joint basis vector of $(L^2_1, L_{1z})$ and
$(L^2_2, L_{2z})$ by $\left|l_1,l_2,m_1,m_2\right>$ or $\left|m_1,m_2\right>$,
if there is no ambiguity. 

\section {Clebsch-Gordan Coefficients and Binomial Distribution}

With notation in place we now state and prove the following theorems:
\begin{thm} Let ${\bf L}=L_1+L_2$ be as above and let $l=l_1+l_2$. If
$$\left |L=l/2, M=l/2\right>
=\left |m_1=l_1/2, m_2=l_2/2\right>$$
then\newline
$$\left<m_1=l/2 - k_1, m_2=l_2/2 - k_2|L=l/2, M=l/2-k\right>^2=
\frac{\left(\begin{array}{c} 
l_1\\k_1\end{array}\right)
\left(\begin{array}{c}l_2\\k_2\end{array}\right)}{\left(\begin{array}{c}l\\k
\end{array}\right)}$$ 
where $k=k_1+k_2$.\end{thm}
{\bf Proof:} First note that if the operator $L^-$ is applied k times to
$\left |L=1/2, M=1/2 \right>$ we get from equation (5)
\begin{eqnarray} (L^-)^k 
\left |L=l/2, M=l/2\right>
&=&\left [(\frac{l}{2} + \frac{l}{2})(\frac{l}{2} - \frac{l}
{2}+1)\right]^{1/2}
(L^-)^{k-1}\left|\frac{l}{2}, \frac{l}{2}-1\right>\\
&=&[l(l-1)(l-2)\dots (l-k+1)k!]^{\frac 12}\left|\frac l2,\frac l2 - k\right>\\
&=&\left(\begin{array}{c}l\\k\end{array}\right)^{\frac 12}k!\left|\frac l2,\frac l2 - k\right>
\end{eqnarray}  
But also $(L^-)^k=(L_x-iL_y)^k=\sum^{k}_{k_1=0}
\left(\begin{array}{c}k\\k_1\end{array}\right)L^{k_1}_xL^{k_2}_y$.
And applying each term $L^{k_1}_xL^{k_2}_y$ to 
$\left |m_1=l_1/2, m_2=l_2/2 \right>$ gives
\begin{eqnarray*}L^{k_1}_xL^{k_2}_y\left |m_1=l_1/2, m_2=l_2/2\right>&=&
\left(\begin{array}{c}l_1\\k_1\end{array}\right)^{\frac 12}k_1!
\left(\begin{array}{c}l_2\\k_2\end{array}\right)^{\frac 12}k_2!
\left|\frac {l_1}{2}-k_1,\frac {l_2}{2} - k_2\right>\\
&=&\frac{k!\left(\begin{array}{c}l_1\\k_1\end{array}\right)^{\frac 12}
\left(\begin{array}{c}l_2\\k_2\end{array}\right)^{\frac 12}}
{\left(\begin{array}{c}k\\k_1\end{array}\right)}
\left|\frac {l_1}{2}-k_1,\frac {l_2}{2} - k_2\right>
\end{eqnarray*}
It now follows that
\begin{eqnarray*}
\left(\begin{array}{c}k\\k_1\end{array}\right)
L^{k_1}_xL^{k_2}_y\left |m_1=l_1/2, m_2=l_2/2\right>&=&
k!\left(\begin{array}{c}l_1\\k_1\end{array}\right)^{\frac 12}
\left(\begin{array}{c}l_2\\k_2\end{array}\right)^{\frac 12}
\left|\frac {l_1}{2}-k_1,\frac {l_2}{2} - k_2\right>
\end{eqnarray*}
and equating this with equation (8) gives
\begin{eqnarray} 
\left|\frac l2,\frac l2 - k\right>&=&\sum_{k_1=0}^{k}
\frac{\left(\begin{array}{c}l_1\\k_1\end{array}\right)^{\frac 12}
\left(\begin{array}{c}l_2\\k_2\end{array}\right)^{\frac 12}}
{\left(\begin{array}{c}l\\k\end{array}\right)^{\frac 12}}
\left|\frac {l_1}{2}-k_1,\frac {l_2}{2} - k_2\right>
\end{eqnarray}  
In particular,
\begin{eqnarray}
\left<l_1/2 - k_1, l_2/2 - k_2|l/2, l/2-(k_1+k_2)\right>^2=
\frac{\left(\begin{array}{c} 
l_1\\k_1\end{array}\right)
\left(\begin{array}{c}l_2\\k_2\end{array}\right)}{\left(\begin{array}{c}l\\k
\end{array}\right)}
\end{eqnarray}
which is the required result. Note also that the above formula is nothing more
than a hypergeometric distribution.
\begin{thm} If $K_1$, $K_2$ are independent binomial random variables with
distributions ${\it B}(l_1,p)$ and ${\it B}(l_2,p)$ respectively and
$M_i=l_i/2-K_i$, for each $i$ then\newline 
(i) $K=K_1+K_2$ will have ${\it B}(l_1+l_2, p)$ distribution\newline  
(ii) $P(M=\frac{l}{2}-k)=P(K=k)=
\left(\begin{array}{c}l\\k
\end{array}\right)p^k(1-p)^{l-k}$\newline 
\begin{eqnarray*}(iii)\ P(M_1=\frac{l_1}{2}-k_1, M_2&=&\frac{l_2}{2}-k_2
|M=\frac{l}{2}-k)\\&=&
\left<M_1=l/2 - k_1, M_2=l_2/2 - k_2|L=l/2, M=l/2-k\right>^2
\end{eqnarray*}
where $k=k_1+k_2$. 
\end{thm}
{\bf Proof:}(i) It is a well known result in probability theory that the sum of 
two independent binomial random variables with common parameter $p$ is itself
a binomial random variable with parameter $p$. \cite{bd} Indeed, since $K_1$ 
and $K_2$
are binomial r.v.'s with moment generating functions $[pe^t+(1-p)]^{l_1}$ and
$[pe^t+(1-p)]^{l_2}$ respectively, then the moment generating function 
of $K=K_1+K_2$ is
$[pe^t+(1-p)]^{l_1+l_2}$ which means $K$ is a binomial random variable with
binomial distribution ${\it B}(p, l_1+l_2)$.\newline 
(ii) $P(M=l_i/2-k)=P(l_i/2-K=l/2-k)=P(K=k)=
\left(\begin{array}{c}l\\k
\end{array}\right)p^k(1-p)^{l-k}$\newline 
by definition of binomial\newline 
(iii) Direct calculation now gives: 
\begin{eqnarray}&&P(M_1=\frac{l_1}{2}-k_1, M_2=\frac{l_2}{2}-k_2
|M=\frac{l}{2}-k)\\
&=&\frac{P(M_1=\frac{l_1}{2}-k_1, M_2=\frac{l_2}{2}-k_2)}{P(M=\frac{l}{2}-k)}\\
&=&\frac{\left(\begin{array}{c} 
l_1\\k_1\end{array}\right)p^{k_1}(1-p)^{l_1-k_1}
\left(\begin{array}{c}l_2\\k_2\end{array}\right)p^{k_2}(1-p)^{l_2-k_2}}
{\left(\begin{array}{c}l\\k
\end{array}\right)p^k(1-p)^{l-k}}\\ 
&=&\frac{\left(\begin{array}{c} 
l_1\\k_1\end{array}\right)
\left(\begin{array}{c}l_2\\k_2\end{array}\right)}
{\left(\begin{array}{c}l\\k
\end{array}\right)}\\ 
&=&\left<M_1=l_1/2 - k_1, M_2=l_2/2 - k_2|L=l/2, M=l/2-k\right>^2
\end{eqnarray}
by Theorem 1. The result follows.

\section{Application}

We now apply the above theorem to calculate the C-G coefficients for a pair of 
spin-1 particles. However, we will also find that it reveals interesting 
information about the probability 
weightings associated with the $\left|1\right>$, $\left|0\right>$,
$\left|-1\right>$ states of an individual particle composing the pair. 
First note that direct calculation using "ladder" operators gives:
\begin{eqnarray}
\left|2,2\right>&=&\left|1,1\right>\\
\left|2,1\right>&=&\frac{1}{\sqrt 2}\left|1,0\right>+
\frac{1}{\sqrt 2}\left|0,1\right>\\
\left|2,0\right>&=&\sqrt{\frac{2}{3}}\left|0,0\right>+\frac{1}{\sqrt 6}
\left|1,-1\right>+\frac{1}{\sqrt 6}\left|-1,1\right>\\
\left|2,-1\right>&=&\frac{1}{\sqrt 2}\left|-1,0\right>+
\frac{1}{\sqrt 2}\left|0,-1\right>\\
\left|2,-2\right>&=&\left|-1,-1\right>.
\end{eqnarray}
We now calculate some of the same coefficients using the above theorems. 
Note that the conditions of Theorem 1 are met, in the sense that
$\left|2,2\right>=\left|1,1\right>$. Hence, the formula can be directly 
applied. Indeed, for $\left|2,0\right>=\left|4/2, 4/2-2\right>$, 
the formula gives 
$$\left<m_1=2/2 - 1, m_2=2/2 - 1|L=4/2, M=4/2-2\right>^2=
\frac{\left(\begin{array}{c} 
2\\1\end{array}\right)
\left(\begin{array}{c}2\\1\end{array}\right)}{\left(\begin{array}{c}4\\2
\end{array}\right)}=\frac{2}{3},$$ 
$$\left<m_1=2/2 - 0, m_2=2/2 - 2|L=4/2, M=4/2-2\right>^2=
\frac{\left(\begin{array}{c} 
2\\0\end{array}\right)
\left(\begin{array}{c}2\\2\end{array}\right)}{\left(\begin{array}{c}4\\2
\end{array}\right)}=\frac{1}{6},$$ 
$$\left<m_1=2/2 - 2, m_2=2/2 - 0|L=4/2, M=4/2-2\right>^2=
\frac{\left(\begin{array}{c} 
2\\2\end{array}\right)
\left(\begin{array}{c}2\\0\end{array}\right)}{\left(\begin{array}{c}4\\2
\end{array}\right)}=\frac{1}{6}$$ 
which clearly correspond to the correct C-G coefficients. 

The same result can also be achieved by applying Theorem 2. However, in
this case, 
the use of conditional probability theory also reveals
unexpected information about the distribution of the spin spectrum of the
spin 1 particle\cite{poh}. 
In turns out, the  C-G coefficients for two spin 1 particles with
$l=l_1+l_2$ can only be derived from conditional probability theory, provided
the spectral distribution of an individual spin 1 particle has a probability
distribution of the form $p^2, 2pq, q^2$, which in the case of $p=q$ becomes
1/4, 1/2, 1/4, in contrast to the current belief of 1/3, 1/3, 1/3. 
Specifically, let $M_i$ where $i=1,2$ be a random variable associated with 
the spin of two independent particles such that
\begin{eqnarray}P(M_i=1)=P(M_i=-1)=\frac{1}{4},\ \ \ 
P(M_i=0)=\frac{1}{2}\end{eqnarray}
Also, let $M=M_1+M_2$ be the sum of their spins and note that $M$ is a random 
variable with values $2, 1, 0, -1, -2$. Then the conditional distribution
\footnote{Recall that for two events $A$ and $B$ defined on a finite sample 
space $S$, the conditional probability of $A$ given $B$ is denoted by $P(A|B)$
and $P(A|B)=P(A\cap B)/P(B)$ provided $P(B)\neq 0$.}
for the state $\left|2,0\right>$ associated with the two independent particles 
gives 
\begin{eqnarray}&P&(M_1=0, M_2=0|M=0)=\frac{2}{3},\\
&P&(M_1=1, M_2=-1|M=0)= 
P(M_1=-1, M_2=1|M=0)=\frac{1}{6}\end{eqnarray}
which coincides with the C-G calculation.
On the other hand, if
\begin{eqnarray}P(M_i=1)=P(M_i=0)=P(M_i=-1)=\frac{1}{3}\end{eqnarray}
then direct calculation gives 
\begin{eqnarray*} P(M_1=0,M_2=0|M=0)&=&P(M_1=1,M_2=-1|M=0)\\
&=&P(M_1=-1,M_2=1|M=0)=\frac{1}{3}\end{eqnarray*}
which are the C-G coefficients associated with the singlet state:
\begin{eqnarray}
\left|0, 0\right>&=&\frac{1}{\sqrt 3}\left|1,-1\right>+
\frac{1}{\sqrt 3}\left|-1,1\right>-\frac{1}{\sqrt 3}\left|0,0\right>.
\end{eqnarray}


\begin{thebibliography}{99}
\bibitem{bd} Bickel and Doksum, {\em Mathematical Statistics: Basic Ideas and
Selected Topics}, 
455-456(1977), Holden-Day.
\bibitem{brink} Brink and Satchler, {\em Angular Momentum},30-35(1968), 
Clarendon Press
\bibitem{poh}Paul O'Hara, {\em The spin-statistics theorem -- did Pauli get
it right}, arXiv:quant-ph/0109137.
\end{thebibliography}
\end{document}